# Superconductivity in quasi-one-dimensional $Cs_2Cr_3As_3$ with large interchain distance

Zhang-Tu Tang[1,3], Jin-Ke Bao[1,3], Zhen Wang[1,3], Hua Bai[1,3], Hao Jiang[1,3], Yi Liu[1,3], Hui-Fei Zhai[1,3], Chun-Mu Feng[1], Zhu-An Xu[1,2,3] and Guang-Han Cao[1,2,3]*

Abstract

Since the discovery of high-temperature superconductivity (SC) in quasi-two-dimensional copper oxides, a few layered compounds, which bear similarities to the cuprates, have also been found to host unconventional SC. Our recent observation of SC at 6.1 K in correlated electron material $K_2Cr_3As_3$ (J. K. Bao et al., arXiv: 1412.0067) represents an obviously different paradigm, primarily because of its quasi-one-dimensional (Q1D) nature. The new material is structurally featured by the $[(Cr_3As_3)^{2-}]_\infty$ double-walled subnano-tubes composed of face-sharing $Cr_{6/2}$ ($As_{6/2}$) octahedron linear chains, which are well separated by columns of $K^+$ counterions. Later, an isostructural superconducting $Rb_2Cr_3As_3$ was synthesized, thus forming a new superconducting family. Here we report the third member, $Cs_2Cr_3As_3$, which possesses the largest interchain distance. SC appears below 2.2 K. Similar to the former two sister compounds, $Cs_2Cr_3As_3$ exhibits a non-Fermi liquid behavior with a linear temperature dependence of resistivity in the normal state, and a high upper critical field beyond the Pauli limit as well, suggesting common unconventional SC in the Q1D Cr-based material.

[1] Department of Physics, Zhejiang University, Hangzhou 310027, China
[2] Stat Key Lab of Silicon Materials, Zhejiang University, Hangzhou 310027, China
[3] Collaborative Innovation Centre of Advanced Microstructures, Nanjing 210093, China
* Corresponding author (email: ghcao@zju.edu.cn)

The revolutionary discovery of high-temperature SC in cuprates in 1980s [1] has inspired unprecedented enthusiasm to explore new superconductors, especially in transition-metal compounds. The continuous efforts over a quarter century lead to discoveries of unconventional SC in a few classes of materials including iron pnictides [2], which bear two fundamental similarities to the cuprates — strong electron correlations and quasi two dimensionality. It is of great interest to explore possible unconventional SC in a Q1D material with significant electron correlations.

Our recent observation of SC at $T_c$ = 6.1 K in $K_2Cr_3As_3$ realized the above possibility [3]. Unconventional SC was preliminarily evidenced by the peculiar properties including linear temperature dependence of resistivity in the normal state and extremely high upper critical field exceeding the Pauli limit by a factor of four [3]. $K_2Cr_3As_3$ crystallizes in a hexagonal lattice, comprising of $[(Cr_3As_3)^{2-}]_\infty$ double-walled subnano-tubes and columns of $K^+$ counterions. Consistently, the first-principles calculations [4] indicate two Q1D Fermi surface sheets in addition to a 3D Fermi surface sheet. The electronic states nearby the Fermi level are dominated by Cr-$3d_{xy}$, $3d_{x^2-y^2}$, and $3d_{z^2}$ orbitals, and the renormalization of electronic density of states is appreciable (by a factor of three), indicating significant electron correlations. Large ferromagnetic spin fluctuations [4] and/or frustrated magnetic fluctuations [5] were predicted by theoretical calculations, and such spin fluctuations were definitely shown by the $^{75}$As nuclear spin-lattice relaxation rate [6]. Accompanied with the strong spin fluctuations, very interestingly, a novel 1D Tomonaga-Luttinger liquid (TLL) [7-11] in the normal state is evidenced by the NQR measurements [6]. So far, all the investigations [3-6] consistently point to unconventional SC in the Q1D Cr-based material.

It is known that Peierls transition easily occurs in Q1D metallic systems [11,12], which prevents the appearance of SC [13]. The interchain coupling serves as a crucial control parameter, thus it is of interest to explore other possible analogues with varying interchain coupling strength. By the replacement of $K^+$ with $Rb^+$, an isostructural compound $Rb_2Cr_3As_3$ was successfully synthesized [14]. Indeed, the interchain distance, measured by the $a$ axis, expands by 3%. $T_c$ is consequently decreased to 4.8 K with broadening of the superconducting transition. In this Letter, we tried to further enlarge the lattice by the element replacement with the largest cation $Cs^+$ (except for $Fr^+$, which is a radionuclide), which gives birth to the third member of the $Cr_3As_3$-chain based family. The interchain distance achieves 10.6 Å, 6% larger than that in $K_2Cr_3As_3$. Nevertheless, SC emerges at 2.2 K.

Polycrystalline samples of $Cs_2Cr_3As_3$ were prepared by solid-state reactions in vacuum, similar to the previous reports [3,14]. Details of the sample preparation are presented in the following **Experimental Section**. Note that the resultant sample is very air sensitive, and exposure to air should be avoided as far as possible.

The powder X-ray diffraction (XRD) pattern of the as-prepared specimen shows no obvious impurities, and it can be well indexed by a hexagonal lattice with unit-cell parameters close to those of $K_2Cr_3As_3$ and $Rb_2Cr_3As_3$. We made a Rietveld analysis, using the code RIETAN-FP[15], based on the $K_2Cr_3As_3$ structural model [3]. The XRD refinement profile is shown in Figure 1, and the resulted crystallographic parameters are tabulated in Table 1. The successful refinement indicates that $Cs_2Cr_3As_3$ is also isostructural to $K_2Cr_3As_3$, but the $a$ axis, or the interchain distance, is 6% larger than that of $K_2Cr_3As_3$ [3]. Meanwhile, the $c$ axis increases only by 0.4%. Surprisingly, the in-plane Cr1−Cr1 distance is even smaller than the Cr2−Cr2 distance, contrary to the cases in

$K_2Cr_3As_3$, and especially in $Rb_2Cr_3As_3$ [14]. The average Cr–Cr bond length, $<d_{Cr-Cr}>$, of $Cs_2Cr_3As_3$ is 2.64 Å, almost identical to that in $K_2Cr_3As_3$. This clearly indicates weakening of the interchain coupling, with the $[(Cr_3As_3)^{2-}]_\infty$ chains hardly changed, by the Cs substitution.

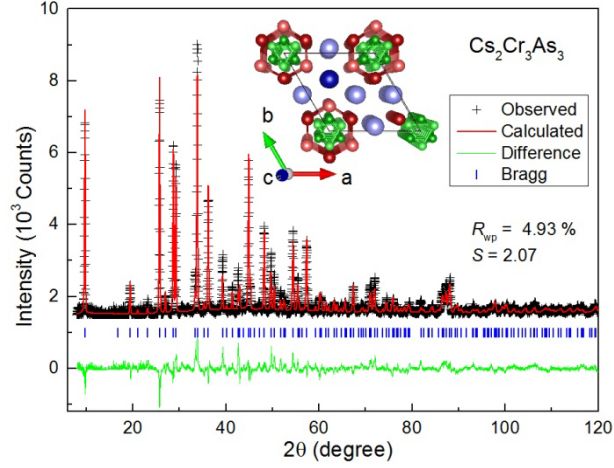

**Figure 1** X-ray diffraction and its Rietveld refinement profile for the $Cs_2Cr_3As_3$ sample. The inset displays the crystal structure which contains $[(Cr_3As_3)^{2-}]_\infty$ double-walled subnano-tubes (DWSTs). The outer wall of the lower-right DWST is stripped off to show the face-sharing $Cr_{6/2}$ octahedron linear chains (in green). $Cs^+$ cations (in blue) well separate the DWSTs.

**Table 1** Experimental crystallographic parameters refined by a Rietveld analysis of the X-ray diffraction data collected at room temperature for $Cs_2Cr_3As_3$.

| Space group | | | P-6 m 2 (#187) | |
|---|---|---|---|---|
| $a$ (Å) | | | 10.605(1) | |
| $c$ (Å) | | | 4.2478(5) | |
| Atom | $x$ | $y$ | $z$ | Wyck. |
| As1 | 0.8399(3) | 0.1601(3) | 0 | 3$j$ |
| As2 | 0.6708(3) | 0.8354(3) | 0.5 | 3$k$ |
| Cr1 | 0.9143(5) | 0.0857(5) | 0.5 | 3$k$ |
| Cr2 | 0.8375(4) | 0.9187(4) | 0 | 3$j$ |
| Cs1 | 0.5328(2) | 0.0655(2) | 0.5 | 3$k$ |
| Cs2 | 1/3 | 2/3 | 0 | 1$c$ |

Figure 2 shows temperature dependence of electrical resistivity, $\rho(T)$, for the $Cs_2Cr_3As_3$ polycrystalline sample. The $\rho(T)$ data show metallic conduction from room temperature down to 2.2 K at which superconductivity emerges. Similar to $K_2Cr_3As_3$ and $Rb_2Cr_3As_3$, the resistivity follows $\rho(T) = \rho_0 + AT$, which indicates a non-Fermi liquid (NFL) behavior. In general, the $T$-linear resistivity can be resulted either from strong spin fluctuations near a quantum critical point [16], or from spin-charge separations in a TLL [17]. It was recently revealed [6] that the nuclear spin-lattice relaxation rate, $1/T_1$, obeys a power law with $1/T_1 \propto T^{0.75}$, which unambiguously evidences a TLL state. Hence the NFL behavior was most probably due to its 1D nature.

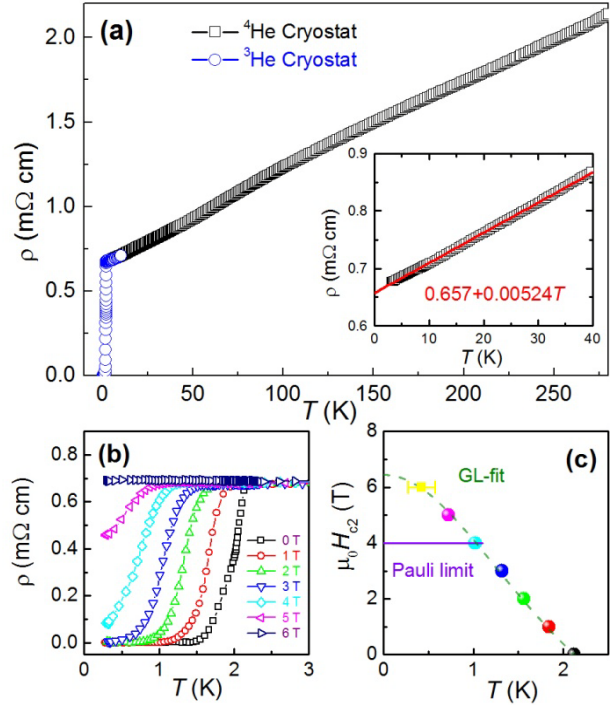

**Figure 2** Temperature dependence of electrical resistivity of the $Cs_2Cr_3As_3$ polycrystalline samples. (a), Full range of $\rho(T)$ data at zero field, with a zoom-in plot in the inset. (b), Superconducting transitions under different magnetic fields, which determine the upper critical fields, $H_{c2}(T)$, as plotted in (c). Note that an onset superconducting transition is detected at 0.44 K under 6 T.

The superconducting transition is explicitly shown in Fig. 2(b), from which the onset-transition temperature $T_c^{onset}$ and the zero-resistance temperature $T_c^{zero}$ was determined to be 2.2 K and 1.5 K, respectively, at zero field. The transition width, defined by the temperature interval with 90% and 10% normal-state resistivity, is 0.43 K, which is quite large relative to the lowered $T_c$. The upper critical fields ($H_{c2}$) were extracted as shown in Fig. 2(c), using the widely used criteria of 90% of the extrapolated normal-state resistivity at the superconducting transition. The data mostly follow the Ginzburg-Landau relation, $H_{c2}(T) = H_{c2}(0)(1-t^2)/(1+t^2)$, where $t = T/T_c$, and the $\mu_0 H_{c2}(0)$ value is then determined to be 6.45 T. Note that the $H_{c2}$ value could be much larger due to the possible anisotropy, if the magnetic field is applied along the **c** axis of the crystal. Even the present $\mu_0 H_{c2}(0)$ value exceeds the Pauli limit ($\mu_0 H_P = 1.84\, T_c \approx 4.0$ T) [18,19], implying the possible spin-triplet spin structure for the superconducting Cooper pairs.

Figure 3(a) shows temperature dependence of magnetic susceptibility ($\chi = M/H$) under an external field of 1 kOe for the bulk polycrystalline sample of $Cs_2Cr_3As_3$. At first sight, the $\chi(T)$ data are Curie-Weiss-like. The Curie-Weiss fit using the general formula $\chi(T) = \chi_0 + C/(T - \theta_P)$ yields a small effective moment of 0.14 $\mu_B$/Cr (other two fitted parameters are: $\chi_0$ = 0.002 emu/mol-fu and $\theta_P = -5.5$ K). The poor fitting as well as the small moment suggests invalidation of Curie-Weiss scenario (the small magnetic moment could be due to magnetic impurities). Indeed, the field dependence of magnetization, $M(H)$, shown in the inset of Fig. 3(a), is actually non-linear even at 150 K or above. The origin of the non linearity needs further investigations using the single crystal samples because of the possible magnetic anisotropy.

Under a low magnetic field, superconducting diamagnetic transition can be seen below $T_c$ = 2.2

K. At the lowest temperature (1.9 K) in our measurement system, the volume fraction of magnetic shielding, measured in the zero-field-cooling mode, is 6%. One may expect a substantially larger value at lower temperatures, because the specific-heat data down to 0.5 K (see below) indicate bulk SC. The inset of Fig. 3(b) shows the $M(H)$ loop at 2 K, which further confirms SC with obvious magnetic flux pinning.

Figure 4 shows the low-temperature specific-heat data for the $Cs_2Cr_3As_3$ polycrystals, plotted with a $C/T$ vs $T^2$ scales. In general, the phonon contribution to specific heat can be well described by the Debye $T^3$ law, thus one expects a good linearity for the $C/T$ vs $T^2$ plot in the normal state. However, deviation from the linearity is obvious even below 5 K. This could be either due to a non-harmonic effect of lattice vibrations or from some unknown sources (e.g., Schottky anomaly from magnetic impurities). If only considering the non-harmonic effect (the possible Schottky anomaly usually contributes a small amount above 2 K), one may fit the normal-state $C(T)$ data by a formula $C = \gamma_n T + \beta T^3 + \delta T^5$. The fitting extracted the electronic specific-heat coefficient of $\gamma_n = 39$ mJ K$^{-2}$ mol-fu$^{-1}$. The $\gamma_n$ value is only one half of that in $K_2Cr_3As_3$ [3] and two thirds of that in $Rb_2Cr_3As_3$ [14]. From the fitted $\beta$ value of 3.8 mJ K$^{-4}$ mol-fu$^{-1}$, the Debye temperature is calculated to be 160 K, which is reasonably smaller than the counterparts of $K_2Cr_3As_3$ (215 K) [3] and $Rb_2Cr_3As_3$ (181 K) [14] because of the heavy cesium incorporated.

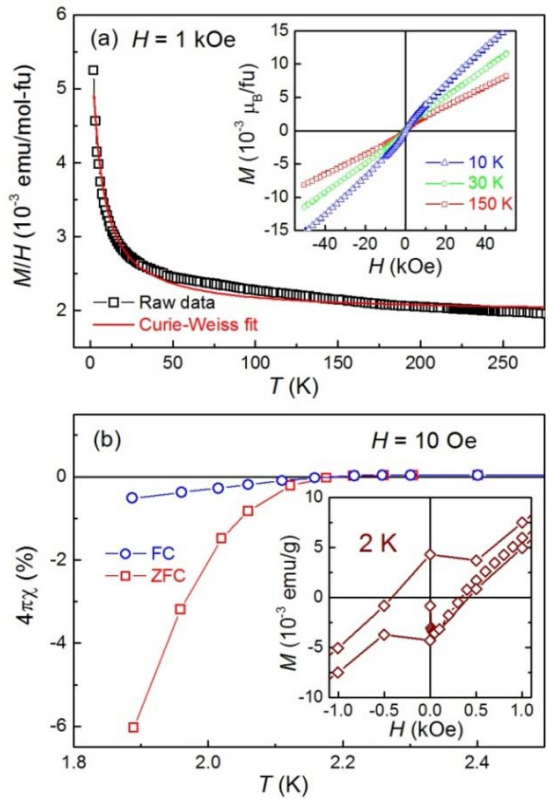

**Figure 3** Magnetic measurements on $Cs_2Cr_3As_3$. (a), Temperature dependence of magnetic susceptibility under a magnetic field of 1 kOe. The inset shows the field dependence of magnetization at three fixed temperatures. (b), Low-field magnetic susceptibility showing the superconducting transition. The inset displays field dependence of magnetization at 2 K. The field was applied in the order of 0 → 50 kOe → −50 kOe → 50 kOe.

The electronic contribution to specific heat, $C_e$, may be obtained by subtracting the phonon contribution, if assuming no other contributions. The inset of Fig. 4 plots a normalized dimensionless electronic specific heat ($C_e/\gamma_n T$) as a function of temperature. The specific-heat jump is clearly seen below 2.2 K, confirming the bulk SC in $Cs_2Cr_3As_3$. However, the superconducting transition is rather broadened. The amplitude of the specific-heat jump is remarkably smaller than those of $K_2Cr_3As_3$ (2.0) [3] and $Rb_2Cr_3As_3$ (1.7) [14]. This could be partly due to degradation of sample during the installation of specimen. It is obvious that the entropy does not conserve, which is most probably owing to the Schottky-anomaly contribution that had not been subtracted. The upturn in $C_e/\gamma_n T$ below 1 K agrees with the existence of Schottky anomaly.

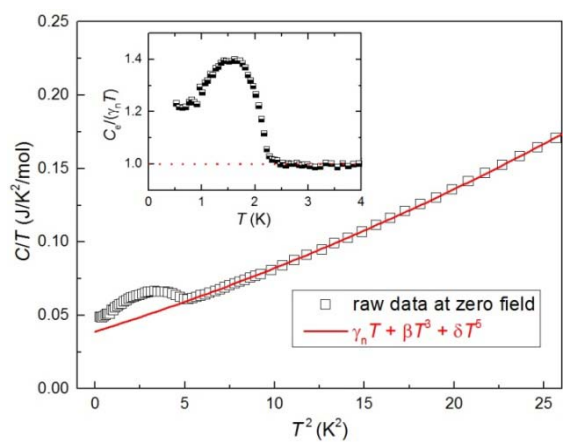

**Figure 4** Low-temperature specific heat data of $Cs_2Cr_3As_3$. See details for the notations and formula in the text.

In summary, we have successfully synthesized a new cesium chromium arsenide, $Cs_2Cr_3As_3$, which belongs to the same family of $K_2Cr_3As_3$ and $Rb_2Cr_3As_3$. Like its sister compounds, $Cs_2Cr_3As_3$ shows similar peculiar physical properties including *T*-linear resistivity and high upper critical field, as measured on the polycrystalline sample, which implies common unconventional SC in the Q1D Cr-based family. Apart from these similarities, $Cs_2Cr_3As_3$ is distinct by the largest interchain distance, corresponding to the weakest interchain coupling. Nonetheless, it still superconducts at 2.2 K. The electronic specific-heat coefficient is only half of that in $K_2Cr_3As_3$. Therefore, the lowered $T_c$ in $Cs_2Cr_3As_3$ could be ascribed to the weakened interchain coupling and/or the reduced density of state at Fermi level. Much work needs to be done to clarify the origin of the possibly unconventional SC.

**Experimental Section**

The $Cs_2Cr_3As_3$ polycrystalline sample was synthesized by a solid state reaction in vacuum using the three elements as the starting materials. First, the starting materials (the elements Cs, Cr and As) all with high purity (≥99.9%, Alfa Aesar) were weighed with 3% excess of Cs in order to compensate the loss during the solid-state reaction. The mixture was loaded in a quartz ampoule, which was evacuated (< $10^{-2}$ Pa) and sealed, followed by slowly heating in a muffle furnace to 423 K for 15 h. After the first stage reaction, the mixture was homogenized, pressed into pellets, put into an alumina tube which was then sealed by arc welding in argon atmosphere in a Ta tube. The sealed Ta tube, jacketed by an evacuated quartz ampoule, was then sintered at 973 K for 24 h. This procedure was repeated to allow a full solid-state reaction. All the operations of weighing, mixing, grinding, pelletizing, etc., were carried out in an argon-filled glovebox with the water and oxygen content below 0.1 ppm.

Powder X-ray diffraction was carried out at room temperature on a PANalytical x-ray diffractometer (Empyrean Series 2) with a monochromatic CuK$\alpha_1$ radiation. We employed Apiezon *N*-grease to protect the sample from being degraded, and the background from the grease was subtracted. The electrical resistivity was measured for the polycrystalline samples, using standard four-terminal method, with a Cryogenic Mini-CFM system using $^4$He cycled refrigerator and on an Oxford superconducting magnet system equipped with $^3$He cryostat. The sample's magnetization was measured on a Quantum Design Magnetic Property Measurement System. The specific-heat capacity was measured by a relaxation technique on a Quantum Design Physical Properties Measurement System with $^3$He option. The heat capacity from the sample holder and grease was deducted.

**Acknowledgements** This work was supported by the Natural Science Foundation of China (Nos. 11190023 and 11474252), the National Basic Research Program (No. 2011CBA00103), and the Fundamental Research Funds for the Central Universities of China.

**Author contributions** Cao GH designed the work and wrote the paper in discussion with Tang ZT, Bao JK, Jiang H and Xu ZA. Tang ZT performed most of the experiments, including the growth, characterizations, and physical property measurements with assistance by Bao JK, Wang Z, Bai H, Liu Y, Zhai HF and Feng CM.

**Conflict of Interest** The authors declare that they have no conflict of interest.